\documentclass{nature}

\usepackage{graphicx}

\usepackage{times}
\usepackage{amsmath}
\usepackage{amssymb}
\usepackage{bm}
\usepackage{physics}
\usepackage{color}
\usepackage{upgreek}

\title{Magnetoconductance modulations due to interlayer tunneling in radial superlattices}

\author{Yu-Jie Zhong$^{1,2}$, Angus Huang$^{3,4}$, Hui Liu$^{5}$, Xuan-Fu Huang$^{1}$, Horng-Tay Jeng$^{3,4,6}$, Jhih-Shih You$^{7}$, Carmine Ortix$^{8,9}$, Ching-Hao Chang$^{1,2\ast}$}

\begin{document}

\maketitle

\begin{affiliations}
 \item Department of Physics, National Cheng Kung University, Tainan 701, Taiwan
 \item Center for Quantum Frontiers of Research \& Technology (QFort), National Cheng Kung University, Tainan 701, Taiwan
 \item Department of Physics, National Tsing Hua University, Hsinchu 30013, Taiwan
 \item Center for Quantum Technology, National Tsing Hua University, Hsinchu 30013, Taiwan
 \item IFW Dresden and W{\"u}rzburg-Dresden Cluster of Excellence ct.qmat, Helmholtzstrasse 20, 01069 Dresden, Germany
 \item Institute of Physics, Academia Sinica, Taipei 11529, Taiwan
 \item Department of Physics, National Taiwan Normal University, Taipei 11677, Taiwan
 \item Institute for Theoretical Physics, Center for Extreme Matter and Emergent Phenomena, Utrecht University, Princetonplein 5, NL-3584 CC Utrecht, Netherlands
 \item Dipartimento di Fisica ``E. R. Caianiello'', Universit\'a di Salerno, IT-84084 Fisciano, Italy
\end{affiliations}

\noindent $^\ast$To whom correspondence should be addressed; E-mail:  cutygo@phys.ncku.edu.tw.\\

\begin{abstract}

Radial superlattices are nanostructured materials obtained by rolling-up thin solid films into spiral-like tubular structures. The formation of these ``high-order" superlattices from two-dimensional crystals or ultrathin films is expected to result in a transition of transport characteristics from two-dimensional to one-dimensional. Here, we show that a transport hallmark of  radial superlattices is the appearance of magnetoconductance modulations in the presence of externally applied axial magnetic fields. This phenomenon critically relies on electronic interlayer tunneling processes that activates an unconventional Aharonov-Bohm-like effect. Using a combination of density functional theory calculations and low-energy continuum models, we determine the electronic states of a paradigmatic single-material radial superlattice -- a two-winding carbon nanoscroll --  and indeed show momentum-dependent oscillations of the magnetic states in axial configuration, which we demonstrate to be entirely due to hopping between the two windings of the spiral-shaped scroll.
 
\end{abstract}

Since the pioneering work of Esaki and Tsu back in the 1970 \cite{esaki1970}, superlattices have introduced a new paradigm for the synthesis of artificial nanoscale material structures with tailored electronic properties. The synthesis of moir\'e superlattices of two-dimensional van der Waals materials, obtained for instance by placing single-layer graphene on aligned hexagonal boron-nitride substrates \cite{tang2013, roth2013,tang2015,Dean2010Oct,Woods2014Jun}, has led to the experimental observation of superlattice minibands forming the well-known Hofstadter butterfly\cite{hof1976} and the related generation of Dirac cone replicas \cite{pon2013,dean2013,hunt2013,Yankowitz2012May,Ortix2012Aug,Wallbank2013Jun}. Likewise, the moir\'e superlattice of magic-angle twisted bilayer graphene has been to shown to yield flat bands associated with emerging correlated insulating  behavior and superconductivity \cite{Cao2018Apr_1, Cao2018Apr_2}. 
Rolled-up nanotechnology \cite{Schmidt2001Mar,Xu2019Jan,Chen2016Jan,Deneke2009Apr} -- a strain-induced technique able to tune planar ultrathin films into complex three-dimensional nanoarchitectures -- provides yet another route to superlattices.  
The rolling-up mechanism can be applied to a huge variety of materials, including metals, insulators, polymers and traditional semiconductor families. 
Functional properties such as the thermal conductivity have been shown to be geometrically tailored in silicon radial superlattices \cite{Li2017Aug}.
Very recently, 
the rolled-up nanotechnology
has been also applied to two-dimensional van-der-Waals materials to create high-order van der Waals superlattices \cite{Zhao2021Mar}. Independent of the material at hand, the preparation of these material structures is expected to modulate, via the effective change of dimensionality, their electronic properties, thereby leading to unconventional transport behavior. For instance, a distinctive property of tubular structures with spiral-like cross sections is that they break the rotational symmetry in the embedding three-dimensional space. As a result, their longitudinal magnetoresistance can exhibit a marked directional dependence \cite{Chang2014Nov}. This angle-dependent magnetoresistance has been predicted to occur in the ballistic regime when considering  rolled-up nanotubes made out of conventional semiconducting materials. The same effect is expected to appear\cite{Chang2017May} , however, also in the diffusive regime characterizing transport in, {\it e.g.}, carbon nanoscrolls: spirally wrapped graphite layers \cite{Xie2009Jul}.  The report of an angle-dependent magnetoresistance behavior in SnS$_2$/WSe$_2$ rolled-ups has provided experimental evidence for the occurrence of this geometry-induced effect \cite{Zhao2021Mar} .  
The aim of this work is to show that another transport hallmark of radial superlattices is the appearance of magnetoconductance modulations in the presence of externally applied axial magnetic fields. 

When subject to axial magnetic fields tubular structures with cross-sections forming closed loops display the well-known Aharonov-Bohm (AB) effect~\cite{Aharonov1959Aug}, which leaves distinctive fingerprints. Aharonov-Bohm (AB) magnetoresistance oscillations have been reported in multiwalled carbon nanotubes (CNT)~\cite{Bachtold1999Feb}. 
Furthermore, the AB effect has been predicted to induce a semimetal to semiconductor transition in single-walled CNTs \cite{Tian1994Feb,Ando2006Apr}. 
Distinctive signatures of the structural symmetry of semiconducting core-shell nanowires can be also inferred from the AB oscillations of the energy levels~\cite{Ferrari2009Apr}. Finally, characteristic fingerprints of the anomalous metallic states in three-dimensional topological insulator nanowires~\cite{Bardarson2010Oct} can be also identified from their associated AB effect.

The open cross section of a radial superlattice prevents a full phase interference between electronic waves traveling in opposite direction at the origin of the AB effect. 
Any magnetoconductance oscillations in an axial configuration should be therefore completely prevented. Contrary to such expectations, we will instead show that radial superlattices can and \emph{do} display magnetoconductance oscillations when an axial magnetic field is externally applied.
Using a combination of density functional theory and an effective continuum low-energy model, 
we will theoretically prove that a carbon nanoscroll (CNS) -- a prototypical radial superlattice -- displays oscillations of the energy levels when subject to an axial magnetic field even in a minimal two-winding geometry.
This effect is entirely due to the tunneling between the two consecutive windings of the spiral structure as proved by the fact that the energy level oscillations disappear by artificially switching off the interlayer coupling. We furthermore prove that the amplitude of the energy level oscillations is strongly dependent upon the momentum in the translationally invariant axial direction of the scroll.

\begin{figure}[tb]
\includegraphics[scale=0.5]{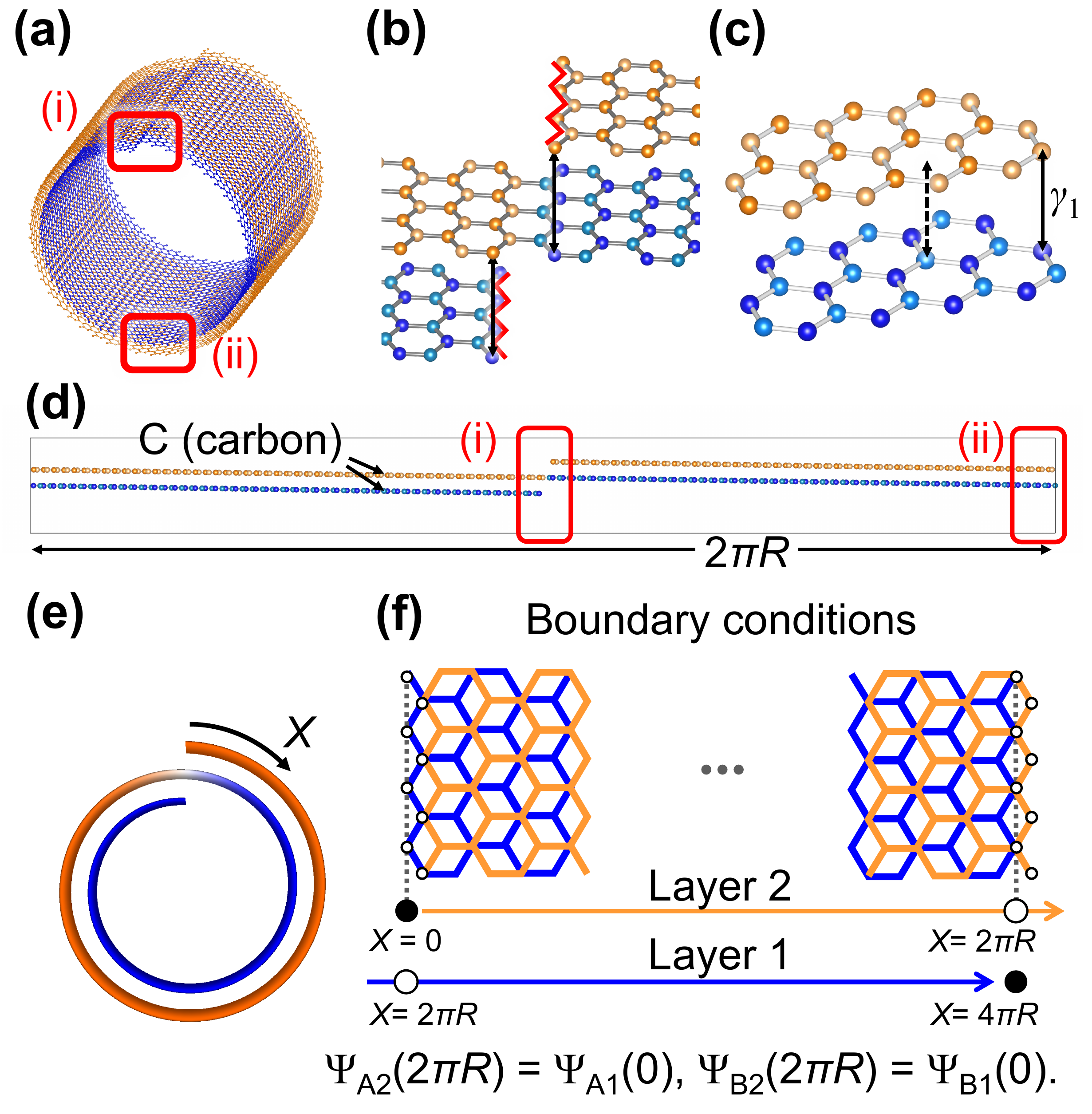}
\caption{ 
 The structure and boundary conditions of two-winding CNSs. (a)  
 The inner layer, layer 1, (outside layer, layer 2) of CNSs   
 is illustrated by blue (orange) line. (b)  
 Zoom in on the region (i) in Fig.~\ref{gns_fig_1}(a).  
 Layer 1 and layer 2 present a zigzag edge, highlighted by red polygonal line.  
 The edges are separated by the connecting region of layer 1 and layer 2. The black arrows are  
 the relative positions of carbon ions. (c) Zoom in on  
 the region (ii) in Fig.~\ref{gns_fig_1}(a). The interlayer coupling 
 is  $\gamma_{1}$. (d) The unit cell, the flattened geometry of CNSs in Fig.~\ref{gns_fig_1}(a), 
has the same topological structure as CNSs. 
  (e) The cross-section with arc length X, and (f) the boundary conditions for the continuum model of two-winding CNSs are shown.  Atom A1 (A2) and B1 (B2) are on the first (second) layer.
}
\label{gns_fig_1}
\end{figure}

We start out by illustrating the mechanism responsible for the onset of magnetoconductance oscillations in a spiralling two-winding carbon nanoscroll. Its geometry is explicitly shown in Fig.~\ref{gns_fig_1}(a). Note that the axial, translationally invariant direction, corresponds to a zigzag direction of the honeycomb lattice. As sketched in Figs.~\ref{gns_fig_1}(b),(c) the interlayer structure is equivalent to the Bernal AB stacking in bilayer graphene.
Therefore, the largest hopping amplitude in the structure corresponds to the interlayer hopping between ``dimer" sites [see Fig.~\ref{gns_fig_1}(c)]. Consider now the 
unrolled nanostructure along the azimuthal direction as shown in Fig.~\ref{gns_fig_1}(d). The strong dimerization between the two layers 
implies that 
at each azimuthal angle the electronic wavefunction has a sizable probability amplitude on both the lower and the upper graphene layers. At the open edges explicitly marked in Fig.~\ref{gns_fig_1}(d), on the other hand, the lower and upper graphene layers have to continuously evolve one into the other in order to reconstruct a two-winding scroll. Consequently, the dimerized electronic wavefunction is ``partially" embedded in a closed geometry.  Therefore, and as a result of the interlayer tunneling, the system can react to an electromagnetic potential precisely as in the conventional AB effect.


 To concretely prove the appearance of such unconventional AB effect, we model a two-winding CNS  with the low-energy ${\bf k \cdot p}$ of bilayer graphene 
along the arclength ($X$ axis) of the nanostrucure while concomitantly using
 special boundary conditions that allow for the tunneling between the two 
 layers (see Fig.~\ref{gns_fig_1}(e) and (f)).

Specifically we use the effective four band model for the 2p$_z$ orbitals on the four atomic sites $A1, B1, A2, B2$ of the unit cell. It can be  
written as follows: \cite{McCann2013}
\begin{equation}
H_{b} = 
 \begin{pmatrix}
  0 & v\pi^{\dagger} & 0 & 0 \\
  v\pi & 0 & \gamma_{1} & 0 \\
  0  & \gamma_{1}  & 0 & v\pi^{\dagger}  \\
  0 & 0 & v\pi & 0  
 \end{pmatrix},    \label{4-band-Hamiltonian}
\end{equation}
where the momentum operators 
are 
$\pi=-i\hbar(k+i\xi k_{z})$ and $\pi^{\dagger}=i\hbar(k-i\xi k_{z})$
with $k_z$ along the CNS axis and $k$ along the arclength direction. In addition, $\xi$ is the valley index that distinguishes between the two inequivalent K points in the Brillouin zone.
The Fermi velocity reads as $v=\sqrt{3}a\gamma_{0}/2\hbar$, where $\gamma_0$ is the intralayer hopping amplitude between sites A1 and B1 (or A2 and B2).  
Finally, we explicitly consider  the interlayer coupling between
 the dimer sites A2 and B1 $\gamma_{1}$, as  shown 
 in Fig.~\ref{gns_fig_1}.

The energy $\varepsilon$ for Eq.~(\ref{4-band-Hamiltonian}) is given by
$\hbar vk_{\pm}=\sqrt{\varepsilon^{2}\pm \gamma_{1}\varepsilon-\hbar^{2}v^{2}k_{z}^{2}}$, 
where ${\pm}$ distinguishes between the valence and conduction band.
Furthermore, the left-moving $L$ and right-moving $R$ wave functions 
read as
\begin{equation}
\Phi_{\pm}^{R(L)} = N_{\pm}
 \begin{pmatrix}
  \mp i\hbar v\left[ (-)k_{\pm}-i\xi k_{z}\right]  \\
  \mp  \varepsilon   \\
   \varepsilon     \\
  -i\hbar v\left[ (-)k_{\pm}+i\xi k_{z}\right]  
 \end{pmatrix}e^{ (-)ik_{\pm}X+ik_{z}z},  \label{wavefunctoin-negative}
\end{equation}
where 
$N_{\pm}$ is a normalization constant.

Within our continuum model, the electronic band structure of the two-winding CNS can be obtained by superimposing the boundary conditions sketched in Fig.~\ref{gns_fig_1}(f). They are given by:
\begin{align}
&\Psi_{A2} (2\pi R) =\Psi_{A1}(0)\notag\\
&\Psi_{B2} (2\pi R) =\Psi_{B1}(0)\notag\\
&\Psi_{A1}(2\pi R)=0 \notag\\
&\Psi_{B2}(0)=0,     \label{boundary-condition}
\end{align}
where $R$ is the radius of the CNSs, with the total arclength of each layer given by
$L=2\pi R$.

To fit the boundary conditions Eq.~(\ref{boundary-condition}), we write a 
generic wave function $\Psi$, which is 
 a combination  of the left-moving $L$ and right-moving $R$ wave functions 
 in the conduction and valence bands
  $\Phi_{\pm}^{R(L)}$. 
 In other words, we write the generic wave function as
$\Psi=a\, \Phi_{+}^{L}+b\, \Phi_{-}^{L}+c\, \Phi_{+}^{R}+d\, \Phi_{-}^{R},$
where $a$, $b$, $c$, and $d$ are 
coefficients 
that we fix by imposing the boundary conditions.

We also use the following microscopic tight-binding parameters: 
The lattice constant is $a=2.46 $ \AA \cite{Saito1998Jul}, whereas
the intralayer and interlayer couplings are $\gamma_{0}=3.16$ eV and $\gamma_{1}=0.381$ eV, respectively \cite{Kuzmenko2009Oct,McCann2013}. 
We also take the arclength of each layer $L=50$ nm.
Thus the total arclength of the two-winding nanoscroll is 
$W=100$ nm.

\begin{figure}[tb]
\includegraphics[scale=0.5]{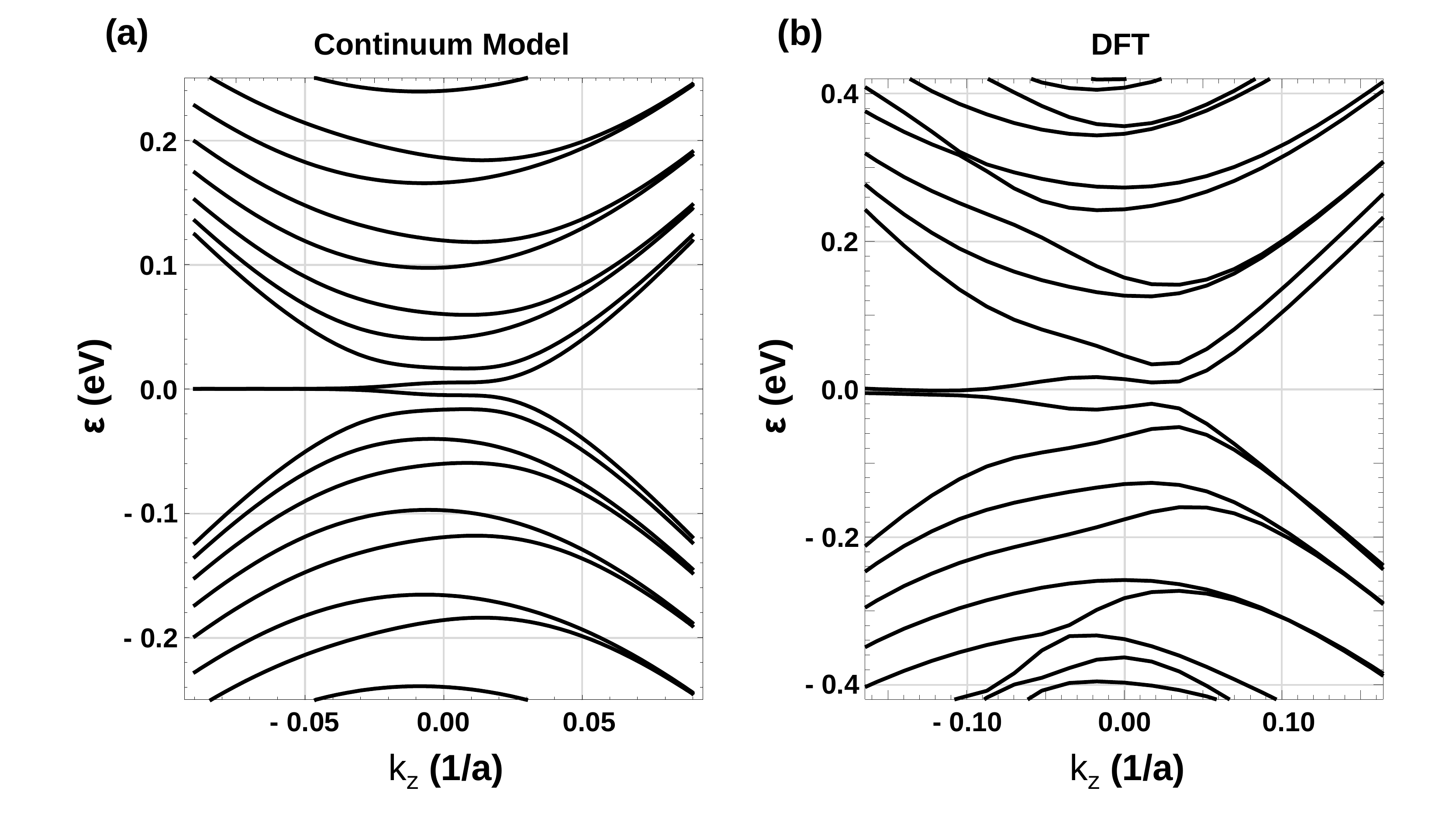}
\caption{ Energy bands of two-winding CNS  for K point. (a) The perimeter is L=50 nm for continuum model and L=20 nm for DFT \cite{fig2b_note2}}.
\label{fig_2b_energyband_0T}
\end{figure}

Fig.~\ref{fig_2b_energyband_0T}(a) shows the energy bands of the two-winding CNSs obtained from our continuum model.
The $k_z=0$ is the projection of the high-symmetry K point, {\it i.e.} the Dirac point in monolayer graphene.  
In perfect analogy with the states found in graphene nanoribbons with zigzag edges, we find that also our two-winding carbon nanoscroll display edge states
for $k_z<0$ momenta [see Fig.~\ref{fig_2b_energyband_0T}(a)].
The density profiles of edge and bulk-like states and their $k_z$-dependent evolutions are provided in Supplementary Information (Section S1).
In order to establish that the continuum model provides an accurate description of the electronic structure of a two-winding CNS, we have thereafter performed density functional theory (DFT) calculations using the structure shown in Fig.~\ref{gns_fig_1}(a) to (d). 
We provide details of the DFT calculations in Supplementary Information (Section S2).
As demonstrated in Fig.~\ref{fig_2b_energyband_0T}, there is
an excellent agreement between the electronic structure obtained using DFT and our effective ${\bf k \cdot p}$ model. 
This consequently allows us to study the effect of an externally applied axial magnetic field using the low-energy model.

To investigate the effect of an externally applied homogeneous axial magnetic field we, as usual, use a minimal coupling with the
vector potential, $\mathbf{A}$, related to the externally applied axial magnetic field
by $\mathbf{B}=\nabla\times \mathbf{A}$.  Since the applied axial magnetic field is considered, $\mathbf{B} \parallel \hat{z}$, 
 we introduce the canonical momenta
$k_{\pm}^{\prime}$
given by
$\hbar k_{\pm}^{\prime}=\hbar (k_{\pm}-\dfrac{eA_X}{\hbar})$.  
The vector potential can be expressed as usual as
$\mathbf{A}=\frac{\Phi}{2\pi}\nabla \theta$, where $\Phi$ is the magnetic flux and $\theta$ is the angle around the cylinder \cite{Ezawa2013Mar}, and more clearly it is given as $A_{X}=\frac{1}{2}BR$ \cite{Chang2014Nov}.
Thus, we have that the wave factors in the wave functions with applied magnetic field must be
written as $e^{ ik_{\pm}^{\prime}X}=e^{ (ik_{\pm} +\frac{2\pi}{L}\frac{\Phi}{ \Phi_0})X}$ and 
$e^{ -ik_{\pm}^{\prime}X}=e^{ (-ik_{\pm} +\frac{2\pi}{L}\frac{\Phi}{ \Phi_0})X}$.  
Here we introduced
the flux quantum is given by $\Phi_0=\frac{2\pi \hbar}{e}$ with $e$ being electronic charge.

\begin{figure}[tb]
\includegraphics[scale=0.5]{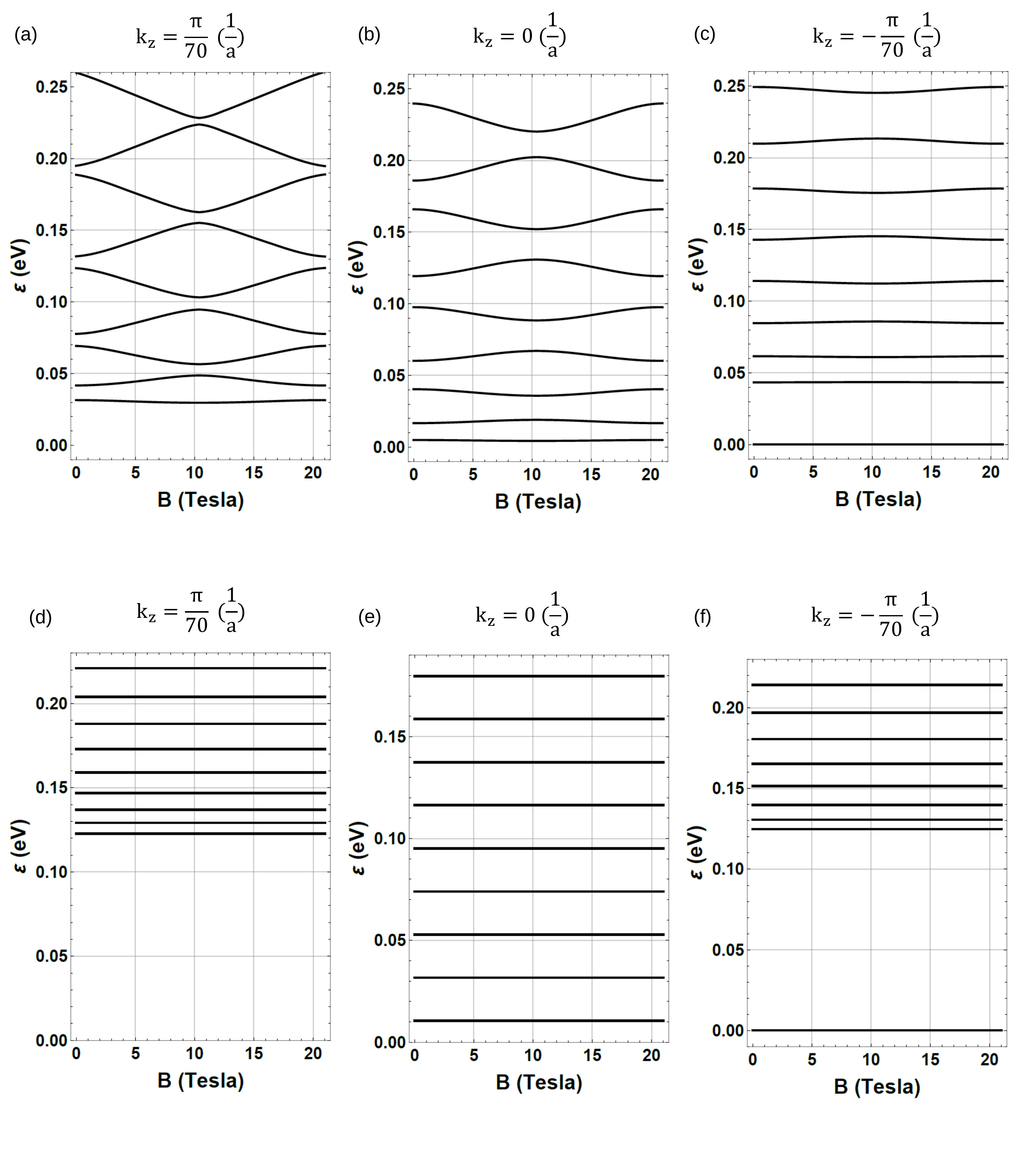}%
\caption{Energy levels $\varepsilon$ vs axial applied magnetic field $B$ at different $k_{z}$. It is shown that Aharonov-Bohm effect presents in two-winding CNSs. (a) to (c) are with
 the interlayer coupling $\gamma_1=0.381$ eV. $\frac{\Phi}{\Phi_0}$ is varied from 0 to 1 with flux quantum: $\Phi_0=4.136\times10^{-15}$ Vs. 
 (d) to (f) are without the interlayer coupling $\gamma_1$.}
\label{fig3_AB_effect_g1_p381}
\end{figure}

Fig.~\ref{fig3_AB_effect_g1_p381} shows our main finding: the 
oscillations of the energy levels as a function of the axial magnetic field intensity 
for different values of the transversal momentum $k_z$ in a structure where the conventional AB effect is precluded.
  Note that the amplitude of the oscillations is enhanced for $k_z>0$ and more attenuated by decreasing the longitudinal momentum toward the $k_z<0$ region.

This result indicates that the 
electronic states with $k_z < 0$ in the nanoscroll are similar in nature to the states in a flat ``unrolled" bilayer graphene nanoribbon 
which, due to the open geometry, are not expected to display energy oscillations.
As proved in the Supplementary Information (Sections S4 and S6) these states without oscillations of the magnetic energy levels satisfy the following criterion on their tube axis momentum
\begin{equation}
\xi k_z\ll k_c(\varepsilon) = -\sqrt{\frac{\varepsilon(\gamma_1-\varepsilon)}{2\hbar^2v^2}}
\label{eq_kc}
\end{equation}
An example of the states satisfying the criterion above is shown in Fig.~3(c). We also emphasize that also the zero energy edge state [see Fig.2] also satisfy this criterion and thus is not practically affected by the electromagnetic potential. We finally note that the criterion in the opposite valley simply interchanges the sign of the transversal momentum $k_z$ as mandated by time-reversal invariance.

In Fig.~\ref{fig3_AB_effect_g1_p381} we also present the magnetic energy states by artificially neglecting the dimer interlayer hopping amplitude $\gamma_1$. The 
magnetic oscillations of the energy levels are completely absent thus demonstrating that the presence of 
our unconventional AB effect in a spiral open structure is entirely due to electronic tunneling between the layers.
%
%

We finally evaluate the consequence of the 
oscillations of the magnetic energy levels in transport.
In particular, we consider the two-terminal conductance in the ballistic regime.   
We therefore use the Landauer formula \cite{Landauer1957Jul,Bagwell1989Jul,Chang2014Nov}, 
\begin{equation}
G(E_F,T)=\int_{-\infty}^{\infty}G(E,0)\frac{\partial f}{\partial E_F}dE,
\end{equation}
where $f$ denotes as the Fermi-Dirac distribution function, and $E_F$ is the Fermi energy. As usual, $G(E,0)=e^2 N_s /h $ is the conductance at 
zero temperature, which is simply given by the number of occupied bands in our quasi-one-dimensional nanostructure.  
Note also that we do not account for the small spin-orbit coupling of graphene and therefore the spin simply gives a multiplicity two to the conductance.

\begin{figure}[tb]
\includegraphics[scale=0.5]{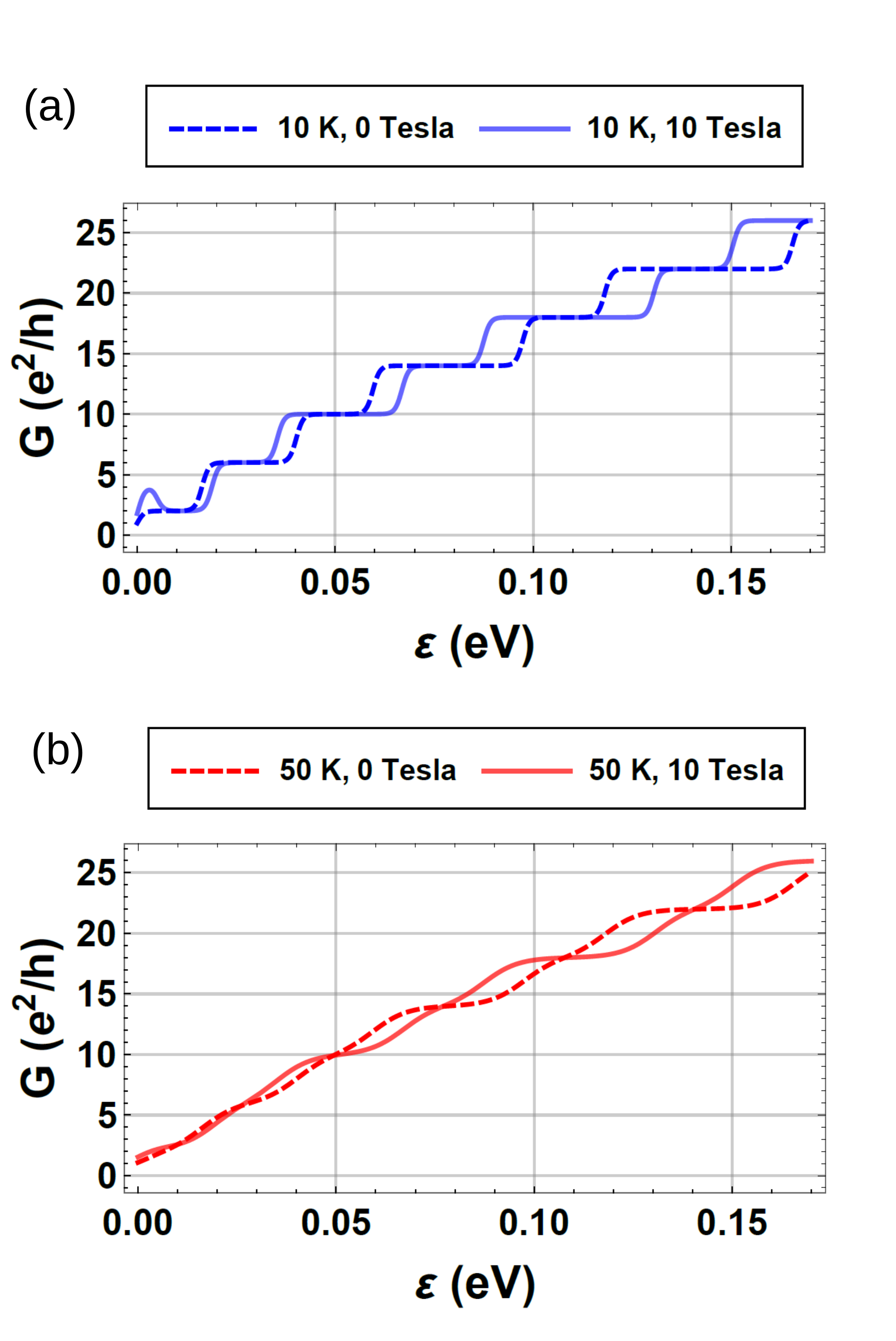}
\caption{The conductance of two-winding carbon nanoscrolls for (a) low temperature (10 K) and (b) high temperature (50 K). Dashed line indicates the applied magnetic field $B = 0$ Tesla, solid line is for $B \approx 10$ Tesla (10.3949 Tesla in the numerical calculation). }
\label{fig_4_conductance}%
\end{figure}

We show the behavior of the conductance in the absence and presence of an axial magnetic field in Fig.~\ref{fig_4_conductance}. 
The oscillations of the magnetic energy levels is reflected in a finite magnetoconductance as evidenced by the change of ballistic conductance as the field strength is varied.
 In Fig.~\ref{fig_4_conductance}(a), low temperature and without applied magnetic field case (10 K and 0 Tesla), the widths of plateaux vary alternatively between large and small. It is similar to the magnetic case; yet for magnetic field case  
 (10 K and 10 Tesla), plateaux change their narrow ones to broad, and vise versa.
 The similar phenomena occur in the high temperature (50 K) 
(Fig.~\ref{fig_4_conductance}(b)) and the medium temperature (25 K) 
(Fig.~S10 in Supplementary Information).
In the low-temperature regime, we have the conventional quantization in units of $2 e^2 / h$ of a quantum point contact \cite{vanWees1988Feb}  with different plateaus as we sweep the Fermi energy 
(see Fig.~\ref{fig_4_conductance}(a)). The difference in the plateaux structure in the presence of the axial magnetic field is the fingerprint of the 
the magnetoconductance modulations.
Note that the axial magnetic field effects on the two terminal conductance persist even in the high temperature regime where the quantization of the conductance is lost due to the thermal smearing, as is shown in Fig.~\ref{fig_4_conductance}(b).

To wrap up, we have considered with a combination of DFT calculations and an effective low-energy ${\bf k \cdot p}$ model, the electronic properties of a two-winding CNSs and established in the presence of an axial magnetic field the occurrence of 
oscillations of the magnetic
energy levels as the intensity of the magnetic field is increased. We have found that because of the presence of an interlayer tunneling
there is an unconventional Aharonov-Bohm-like effect with electrons responding to an electromagnetic potential even if the
structure does not possess a closed cylindrical geometry. In addition, the presence of both bulk and edge states yields a different oscillation amplitude of the energy levels for different values of the momentum along the nanoscroll axis. 
We demonstrate this feature of magnetoconductivity is originated from a unique nature of the nanoscroll -- its states varies between the tube states and the ribbon states when the momentum changes.

This newly established phenomenon is not specific of carbon nanoscrolls. It can indeed appear in different radial superlattices obtained by rolled-up technology including the recently synthesized high-order van der Waals superlattices. 
Moreover, its presence is very robust against moderate disorder (see Section S6 in Supplementary Information).
Since in all these nanostructures electronic tunneling between the various layers is present, we expect the effect we have unveiled in our work to be observable in the experimental realm.

\section*{Acknowledgements} 
\vspace{-7mm}
We acknowledge the financial support by the Ministry of Science and Technology (Grants No. MOST-107-2112-M006-025-MY3, No. MOST-108-2638-M-006-002-MY2,
No. MOST 110-2112-M-006-020-, and No. MOST 110-2112- M-003-008-MY3) and National Center for Theoretical Sciences in Taiwan.
C. O. acknowledges support from a VIDI grant (Project 680-47-543) financed by the Netherlands Organization for Scientific Research (NWO).
C.-H. C. acknowledges support from the Yushan Young Scholar Program under the Ministry of Education (MOE) in Taiwan.
We acknowledge technical assistance by Botsz Huang and You-Ting Huang.
The authors gratefully acknowledge many helpful discussions with Szu-Chao Chen, Hsiu-Chuan Hsu, Ion Cosma Fulga, Chao-Cheng Kaun, Chao-Ping Hsu, and Chong-Der Hu.


\section*{References}
\bibliography{references}




\section*{\large METHODS}
\vspace{-7mm}
\subsection{First-principle calculations.}
To study the two-winding carbon nanoscrolls (CNSs), first-principle simulations based on density functional theory (DFT) are developed to confirm our model matching to the realistic materials. In the first-principle calculations, the Vienna Ab initio Simulation Package (VASP)\cite{Kresse1996Oct,Kresse1993Nov} with the projector augmented wave (PAW) method \cite{Blochl1994Dec,Kresse1999Jan} is used. The Ceperley-Alder (CA) type local density approximation (LDA) \cite{Ceperley1980Aug} is included as exchange-correlation function. 
The energy cut off (k-points grid) with 400 eV ($12 \times 1 \times 1$) is utilized in simulations. 

\subsection{Data availability.} The data that support the findings of this study are available from the corresponding author upon reasonable request.





\end{document}